\documentclass[aps,pre,onecolumn,epsf,graphicx,a4]{revtex4}
\usepackage[dvips]{graphicx}
\usepackage{amsfonts}
\usepackage{epsfig}
\usepackage{amsmath}
\usepackage{amssymb}

\def\pd2x{{\partial^2 \over \partial x^2}}

\usepackage{epsfig,latexsym}

\newcommand \bew {\begin{widetext}}
\newcommand \enw {\end{widetext}}

\begin{document}

\title{\bf\noindent
Fluctuations in the Site Disordered Traveling Salesman Problem}

\author{David S. Dean$^{(1)}$ and David Lancaster$^{(2)}$}

\affiliation{
(1) Laboratoire de Physique Th\'eorique,  UMR CNRS 5152, IRSAMC, Universit\'e 
Paul Sabatier, 118 route de Narbonne, 31062 Toulouse Cedex 04, France\\
(2) Harrow School of Computer Science, University of Westminster, 
Harrow, HA1 3TP, UK
}
\date{24 August 2007}
\begin{abstract}
We extend a previous statistical mechanical treatment of
the traveling salesman problem by defining a 
discrete ``site disordered'' problem in which
fluctuations about saddle points can be computed. 
The results clarify the basis of our original treatment, 
and illuminate but do not resolve
the difficulties of taking the zero temperature limit 
to obtain minimal path lengths.
\end{abstract}

\maketitle
\vspace{.2cm} \pagenumbering{arabic}

\section{Introduction}

Theories describing disordered materials are based either
on disorder arising from randomness in the strength of bonds
or from randomness in site location. Although most theoretical
progress has been made in the bond-disordered case, site-disorder
is probably closer to the picture of real materials. 
A similar situation exists in the best known problem of
combinatorial optimization: the traveling salesman problem (TSP),
here taken to be the stochastic version.
The Euclidean problem is defined in terms of randomly 
assigned city locations, but analytic progress is most
developed for the random-link model \cite{tspps2,tspcav} in which inter-city
distances are chosen independently from some probability distribution.
Recently we have presented a functional statistical mechanical technique 
to compute the average length of the TSP path at finite temperature
for the site-disordered model \cite{short,long}. In this paper we define a
discrete model that  provides a more rigorous foundation for 
our functional approach in the continuum and allows us
to compute fluctuation terms that illuminate the
difficulties of taking the zero temperature limit to obtain
minimal path lengths.

The site disordered TSP is a straightforward generalization of
the ordinary TSP that populates a graph
consisting of $M$ nodes or sites with $N$ cities. 
The graph can be abstract or for geometric applications we 
could take it to be a regular lattice.
Site disorder arises since the number of cities
on each site, which may be 0, 1 or greater than 1, is part of
the specification of the problem, and we will have in mind a quenched
random distribution. This problem is defined 
in the first section below and various regimes depending
on the average density of cities, $N/M$, are identified.
Section \ref{coarse} shows how the results 
of our previous functional formalism
can be obtained from the 
$N\gg M$ high density or ``coarse'' regime of the discrete site disordered TSP
for a fixed rather than random distribution of cities.
These results correspond to the leading order saddle point approach,
and we show how they can be extended by 
analyzing fluctuations in the discrete model.

Our interest in the fluctuations arises from the possibility
that they may be used to probe minimal TSP path lengths.
This was not possible in our previous work since the
leading saddle point term always gives path lengths that
grow proportional to $N$ and correspond to strictly finite
temperature. In certain cases the fluctuation terms have the correct scaling 
and may be expected to dominate in the low temperature
regime where the minimal path lengths can be accessed.
Unfortunately this program is not realized.
The fluctuations can be computed explicitly
for some graphs with simple structure and we discover that 
in the limits needed to find minimal path lengths the
saddle point approximation breaks down in a complicated manner.
The manner of this breakdown is most clearly exposed for the case
$M=2$ that is analyzed in detail in the appendix.

The remaining sections of the paper discuss the site disordered problem 
in two other regimes.
Firstly the case $N=M$, precisely the ordinary TSP, 
where it is interesting
to see that the saddle point approximation breaks
down in a completely different way from in the $N\gg M$ coarse regime.
Secondly the case
$N\ll M$, which for a regular, dilutely occupied lattice
can also correspond to the ordinary geometric TSP,
and we discuss the significant interest of this regime,
and point out the additional techniques that may be needed to resolve 
the problem in this regime.

\section{Formalism}

Consider a graph consisting of 
$M$ sites with a matrix $J_{ij}$ specifying the
distance between any pair of sites $i,j = 0,1,2\dots M-1$. 
For simplicity the diagonal terms $J_{ii}=J_D$ are taken all equal 
but the matrix need neither be symmetric nor positive though
we shall generally take it to be so.
Imagine that there are $n_j$ cities ($n_j \ge 0$) at site $j$.
We will call these quantities the ``occupation numbers'' and
for a total of $N$ cities they obey $\sum_{j=0}^{M-1} n_j = N$. 
A path through the $N$ cities is specified by the sequence of
site indices $j_0,j_1\dots j_{N-1}$, so depending on the 
occupation numbers ($n_j$'s),
some sites might never be visited, or visited more than once.
A particular path,  $j_0,j_1,j_2\dots j_{N-1}$, can also be characterized
by an element of the permutation group, $\sigma \in {\Sigma_N}$, 
consisting of a permutations of $M$ site indices, 
each repeated $n_j$ times, so as to have a total of $N$ indices.
In this notation the path is written
$\sigma_0,\sigma_1\dots \sigma_{N-1}$,
and the cost or total length for this path is:
\begin{equation}
D(\sigma) = \sum_{m=0}^{N-1} J_{\sigma_m,\sigma_{m+1}}
=\sum_{m=0}^{N-1} J_{j_m j_{m+1}}.
\end{equation} 
Where we identify $j_N$ with $j_0$. As is often the case in this kind of 
problem, the cost is independent of
cyclic changes and also (if $J_{ij}$ is symmetric) reversals
of the permutation.

A full statement of the problem involves a specification of the
occupation numbers as well as the cost matrix between all sites.
Because we take a site-disordered approach we shall generally
assume that the cost matrix is a fixed known quantity and 
will often have a geometric basis in mind for which 
the cost matrix follows from a regular lattice graph.
Disorder can appear through the
occupation numbers on each site.

By varying $N/M$, the average number of cities per site, 
we see that this model describes various regimes. 
If we take $N=M$ and set all the $n_k=1$,
then we immediately return to the TSP as usually formulated
with arbitrary cost matrix $J_{ij}$ that may, or may not, have a 
geometric basis. 
When we consider geometric problems such as the Euclidean TSP
 it is numerically advantageous to imagine the sites
as being on a regular lattice, thus allowing the use of 
integer arithmetic to compute $J_{ij}$~\cite{dsjohnson}, and by considering
sufficiently dilute occupation $N\ll M$ the  continuum result 
is recovered.
In this Euclidean lattice case, the $N=M$ situation
is trivial as it corresponds to a fully occupied lattice
where the optimum TSP paths only involve nearest neighbour links.
The approach to the continuum and standard TSP as  lattice occupation is
made progressively more dilute, $N\ll M$,
has been considered numerically in \cite{chakrabati}.
Analytically, the most tractable regime is when $N\gg M$ and this
may be regarded as a coarse, in the renormalisation group sense,
approximation to the TSP in
which groups of close cites are bunched together at the same site
ignoring the detailed distances between them.
Even in this coarse regime, information about the standard TSP
can be obtained by arranging for the diagonal distances $J_{ii}$
to be small.

The statistical mechanical partition function for this problem 
at temperature  $1/\beta$ is defined by 
the following sum over permutations that visit sites
the correct number of times. 
\begin{equation}
Z_N(\beta,\lbrace n_j\rbrace) = 
\sum_{\sigma\in {\Sigma_N}} 
\exp\left( -\beta D(\sigma)\right)
\label{eqpart}
\end{equation}
This partition function $Z_N(\beta,\lbrace n_j\rbrace) $
depends both on occupation numbers and $J_{ij}$'s and therefore
encompasses both bond and site disorder.
The sum over permutations may be written
as an unconstrained sum over all possible paths of $N$ links with a 
Kronecker delta constraint  
to ensure that site $j$ is visited precisely $n_j$ times.
\begin{equation}
Z_N(\beta,\lbrace n_j\rbrace) =
\sum_{j_0,j_1\dots j_{N-1}=0}^{M-1}
\prod_{k=0}^{M-1} n_k! \ 
\delta_{n_k , \sum_{m=0}^{N-1} \delta_{k,j_m}} \
\exp\left(-\beta\sum_{m=0}^{N-1} J_{j_m j_{m+1}}\right).
\end{equation}
The constraints appearing here should not be confused with the 
constraints that forbid disconnected paths in the 
linear programming approach to the TSP; here the path is
always connected, but must be forced to visit sites the correct
number of times. 
The combinatorial factors of $n_k!$ arise from the different possible
orderings of visiting multiple cities at the same site, 
and are necessary to ensure
that with no weights, $\beta=0$, we recover $Z_N = N!$, corresponding 
to the total number of possible orderings of $N$ cities.
It is implicit in the formalism that $\sum n_k = N$ since this
constraint is enforced by one of the Kronecker deltas.

Writing each Kronecker delta in terms of an integral representation we
find,
\begin{eqnarray}
Z_N(\beta,\lbrace n_j\rbrace)  &=&
\sum_{j_0,j_1\dots j_{N-1}=0}^{M-1}
\oint \prod_{k=0}^{M-1}  
{dz_k z_k^{n_k -1}n_k!\over 2\pi i}
 \prod_{m=0}^{N-1}{1\over z_{j_m}}  \
\exp\left(-\beta\sum_{m=0}^{N-1} J_{j_m j_{m+1}}\right)\\
&=&
\sum_{j_0,j_1\dots j_{N-1}=0}^{M-1}
\int_{-\pi}^{\pi} \prod_{k=0}^{M-1}  
{d\mu_k n_k!\over 2\pi} \,
\exp\left(i\sum_{k'=0}^{M-1} n_{k'} \mu_{k'} -i\sum_{m=0}^{N-1} \mu_{j_m} \right)
\exp\left(-\beta\sum_{m=0}^{N-1} J_{j_m j_{m+1}}\right).
\end{eqnarray}
Where the second form is obtained from the contour integral on
the first line by setting $z_k = e^{i\mu_k}$. Although the integrals
are finally evaluated using this second form, it is helpful to keep
the first form in mind to check what deformations of the
contour are possible in the presence of branch cuts or poles.
Rearranging, the partition function can be written in terms of an integral 
over a subsidiary problem,
\begin{equation}
Z_N(\beta,\lbrace n_j\rbrace)  =
\int_{-\pi}^{\pi} 
\prod_{k=0}^{M-1} {d\mu_k n_k!\over 2\pi} 
\exp\left(i\sum_{k'=0}^{M-1} n_{k'} \mu_{k'}\right)
{\cal Z}_N(\beta,\lbrace \mu_j\rbrace ) 
\end{equation}

The subsidiary statistical mechanical problem does not depend on the 
occupation numbers, $n_k$, and is defined as
\begin{equation}
{\cal Z}_N(\beta,\lbrace \mu_j\rbrace ) =
\sum_{j_0,j_1\dots j_{N-1}=0}^{M-1}
\exp\left(-i\sum_{m=0}^{N-1} \mu_{j_m} 
-\beta\sum_{m=0}^{N-1} J_{j_m j_{m+1}}\right)
= Tr\, {\bf T}^N.
\end{equation}
Where the $M\times M$, transfer matrix, ${\bf T}$ is the symmetrized
form
\begin{equation}
{\bf T}= {\bf R B R}.
\end{equation}
${\bf R}$ is the diagonal matrix with diagonal elements
$R_k =  z_k^{-1/2}=\exp(-i\mu_{k}/2)$, and ${\bf B} = \exp(-\beta {\bf J})$
is defined in terms of the distance matrix.
In this paper we shall only consider symmetric distance functions, so the
transfer matrix is symmetric, though generally complex and 
therefore not hermitian.

Distance matrices that have some kind of regular geometric basis
yield additional structure for ${\bf B}$.
In one dimension this is most obvious, for example
a regular one dimensional lattice with 
$J_{ij} \propto |i-j|$ makes ${\bf B}$ Toeplitz, 
and if the lattice has periodic boundary conditions ${\bf B}$
is also circulant. 
For regular lattices in two dimensions 
${\bf B}$ is block circulant with each of the blocks Toeplitz,
or themselves circulant in the case of periodic boundary conditions.
In higher dimensions there is a hierarchical structure of ${\bf B}$ block 
circulant with blocks that are themselves block circulant.
For all these regular lattices the
eigenvectors are related to discrete Fourier transforms \cite{circulant}
We also consider fully connected $M$-dimensional graphs
with  $J_{ij} = J_N + (J_D-J_N)\delta_{ij}$, which are particularly tractable.
We often allow a diagonal term in the cost matrix, as this can
be used to probe certain aspects of the theory. 
We will not consider asymmetric cases in this paper, but
examples such as the wallpaper problem \cite{wallpaper} with 
$J_{ij} = \theta(i-j)$ can be treated.

The solution of the subsidiary problem can be written in terms of 
the eigenvalues $\lambda_a$  of ${\bf T}$,
\begin{equation}
{\cal Z}_N(\beta,\lbrace \mu_j\rbrace ) =
Tr\, {\bf T}^N
=\sum_{p=0}^{M-1} \lambda_p^N
\label{trace}
\end{equation}
For general values of the integration variables $z_k$ or $\mu_k$,
the eigenvalues will be complex.
Under the scaling  $z_k \to a z_k$, 
eigenvalues transform as $\lambda \to \lambda/a$,
and below we shall find that this scaling symmetry
is related to the Lagrange multiplier mode
enforcing the constraint $\sum n_k = N$.

To close this section on the formalism, let us interpret it in  the case of
the $N=M$ regime with all  occupation numbers $n_k=1$, corresponding to the
usual TSP. The 
partition function then becomes,
\begin{equation}
Z_N(\beta)   =
\oint  \prod_{k=0}^{N-1}  {dz_k\over 2\pi i  } \
Tr({\bf T}^N)
\end{equation}
With ${\bf T}={\bf RBR}$ and 
${\bf R}^2 = diag(z_0^{-1},z_1^{-1}\dots z_{N-1}^{-1})$,
the trace is a polynomial in the variables $z_j^{-1}$ 
that acts as a generating function for paths that visit
site $j$ a number of times given by the power of $z_j^{-1}$. 
The weight associated
with a particular path is the exponential of the length of the path.
Since  all paths visit $N$ sites, every term in the 
generating function polynomial has equal degree $N$. 
Depending on the  symmetry of the distance matrix,
the generating function often has some symmetry in the variables $z_j$.
The integration defining the partition function for the TSP 
picks out the pole corresponding to the term
$\prod z_j^{-1}$ in the generating function, thereby requiring each 
site to be visited exactly once. 
From this point of view, the approach is similar
to other combinatorial problems where the generating function
is known, and in many  graph theoretic problems
a similar integration can be used to select the desired term.
For example in the mathematical literature, 
the saddle point technique has been used to obtain
asymptotic formulae  for the number of Eulerian oriented graphs 
\cite{BrendanMcKay}.
For $\beta=0$, distances are irrelevant and vertices are 
either connected or not, so the system becomes purely topological
and the problem is to count Hamiltonian
paths for a graph with incidence matrix $\bf B$.
The formalism also resembles the field theory for studying
random Euclidean matrices \cite{Zee}.


\section{Coarse TSP Regime: $N\gg M$}
\label{coarse}

Consider the limit of many cities, $N\to \infty$, with $M$ fixed.
In this limit the occupancies, $n_k$, scale with $N$, and it is
convenient to rewrite them in terms of a ``density of cities'',
$\rho_k$ where:
\begin{equation}
n_k = {N\over M} \rho_k.
\end{equation}
The normalization of $\rho_k$ is $\sum_{k=0}^{M-1} \rho_k = M$.

In a geometric context, the formalism is now based on densities of 
cities rather than their explicit locations and this
limit might be regarded as viewing the system
at coarse resolution, smearing out the precise city locations.
Such coarsening techniques are indeed used in heuristics for solving 
TSP instances, although in a different way,
for example the Held Korp heuristic \cite{HeldKorp} patches together 
paths from different cells. 

In this regime a diagonal term $J_{ii} = J_D$ in the cost matrix can have a
special significance. If this self connection corresponds to a
distance larger than typical neighbour distances, then it
will have little effect at low temperature and the optimum path for uniformly
distributed cities may correspond to the optimum of the 
standard TSP, repeated $N/M$ times. If on the other hand, 
$J_D$ is short in comparison with other distances,
then the favoured paths at low temperature will satisfy the
requirement to visit cities $n_k$ times by simply looping
that many times before proceeding to a neighbour. In this
case the true TSP path is found simply by ignoring the 
contribution from the loops. 
For various simple models with small
$M$, we have considered this small $J_D$ situation explicitly.
At low temperature, by regarding the
neighbour couplings as corrections to  the diagonal part of ${\bf B}$ 
then a perturbative approach can be used to compute the
eigenvalues of ${\bf T}$ and hence the contribution to the partition function
which correctly predicts the true TSP path.
The hope of being able to compute such terms for arbitrary M is a motive
for studying the fluctuations about the saddle point. 
Unfortunately, it appears that the saddle point method, which
is our main tool, generally fails to give  the correct subleading
term in this limit for technical reasons that are
discussed in the appendix and in the sections where particular
models are considered.

By inserting the solution of the subsidiary problem
into the expression for the full partition function, we find,
\begin{eqnarray}
Z_N(\beta,\lbrace n_j\rbrace)   &=&
\exp\left(\sum_{k=0}^{M-1}  \log n_k!\right)
\sum_{p=0}^{M-1} Z^p_N(\beta,\lbrace \rho_j\rbrace) \\
 Z^p_N(\beta,\lbrace \rho_j\rbrace) &=&
\oint   \prod_{k=0}^{M-1}  {dz_k\over 2\pi i z_k} 
\left(\lambda_p \prod_{k=0}^{M-1} z_k^{ \rho_k/M} \right)^N
=
\int_{-\pi}^{\pi}  \prod_{k=0}^{M-1}  {d\mu_k\over 2\pi} 
\exp N\left({i \over M}\sum_{k'=0}^{M-1} \rho_{k'} \mu_{k'}
+\log \lambda_p\right).
\label{coarseZ}
\end{eqnarray}
Where the contributions from each eigenvalue are now distinguished,
at least by a label $p = 0,1,2\dots M-1$. In practice, globally
labeling the eigenvalues of $\bf T$ as the $z_k$ vary is not straightforward.

By writing in terms of $\rho_k$ the $N$ 
scaling in the exponent is exposed that allows 
the integrals for each $Z^p$ to be evaluated 
using a saddle point approximation \cite{saddle}.
For the simplest application of this technique to be valid, 
the eigenvalue must have no
$N$ dependence so we are forbidden to scale the temperature
with $N$ \cite{tspps1}.
The saddle point equations for each $Z^p$
are,
\begin{equation}
-{1\over \lambda_p}
{\partial \lambda_p \over \partial \mu_k}
=i{\rho_k \over M}.
\end{equation}
Now, by taking derivatives of the eigenequation,
${\bf T v}^p= \lambda_p {\bf  v}^p$, noting that
$\partial T_{ij}/\partial \mu_k = -i(\delta_{ki}+ \delta_{kj}) T_{ij}/2$,
recognizing that the left and right eigenvectors are the same due to the
symmetry of the distance matrix and
using normalization, ${\bf v}^p.{\bf v}^p= \sum_k v_k^p v_k^p = 1$, 
we find the general relation,
\begin{equation}
{1\over \lambda_p}
{\partial \lambda_p \over \partial \mu_k}
= -i (v_k^p)^2.
\end{equation}
So the saddle point equation may be rewritten as,
\begin{equation}
(v_k^p)^2
 =
{\rho_k \over M}.
\end{equation}
The right hand side of this saddle point equation does not depend 
on the eigenvalue index $p$.
Ostensibly we may choose signs independently for each component
of the eigenvector to generate distinct solutions for each $Z^p$,
but we shall discover that the role of the signs is not transparent.
The eigenvector is written  $v^p_k = \zeta^p_k\sqrt{\rho_k/M}$,
where the signs $\zeta^p_k=\pm 1$.
Now decomposing the
transfer matrix and writing 
$s^p_k = \zeta^p_ke^{i\mu_k/2}\sqrt{\rho_k\lambda/M}$, 
the signs disappear and the eigenequations become,
\begin{equation}
s^p_k = 
{1\over M}\sum_{k'} e^{-\beta J_{kk'}}{\rho_{k'} \over s^p_{k'}}. 
\label{sequation}
\end{equation}
For all reasonable choices of the distance matrix
this equation has a solution with real positive $s_k$
(to denote this solution we omit the index $p$),
that can be found by a numerically stable iteration
procedure. In many of the explicit models we consider, 
this solution is
in fact independent of the index $k$;
for example, 
in the case of circulant ${\bf J}$ and $\rho_k=1$, this positive solution
is $s_k = s; s^2 = (1/M)\sum_j B_{ij}$. 
For certain particular choices of the 
distance matrix, additional solutions to (\ref{sequation}) may be found.
For all cases considered, these additional solutions,
while not spurious, are less important in the large
$N$ limit than the real positive solution noted above.
These solutions are discussed in the sections below that deal with 
particular models.

In the remainder of this section
we shall only be concerned with the real positive solution  $s_k$.
The location of the saddle points for this solution
in terms of integration variables $z_k$, is given by:
\begin{equation}
\lambda z_k = M {s_k^2\over \rho_k}
\label{speqn}
\end{equation}
Note that all reference to the signs $\zeta^p_k$ has disappeared.  
Evidently, since the right hand side is real and positive, 
there is a saddle point on the real $z_k$ axis 
with $Re\, \mu_k = 0$. 
At this saddle point the transfer matrix is real and
symmetric and has real eigenvalues.
Moreover (provided that the cost matrix is positive)
the Perron Frobenius theorem  assures us that 
the eigenvector $v_k = +\sqrt{\rho_k/M}$,
where $\zeta^p_k=+1$,
corresponds to the maximal eigenvalue. 

Notice that  under the scaling symmetry $z_k \to a z_k$, 
eigenvalues transform as $\lambda \to \lambda/a$, so the 
combination $\lambda z_k$ appearing in (\ref{speqn}) is invariant.
As mentioned above, this scaling
corresponds to the Lagrange multiplier mode involving a simultaneous shift
of all $\mu_k$, that enforces
the constraint $\sum n_k = N$. Since this constraint is implicitly satisfied
by the $\rho_k$ that we specify, there is no need to integrate over it, and 
we may choose to set $\sum \mu_k=0$ or $\prod z_k =1$.
As we shall see explicitly in section IV, the
stationarity condition associated with this mode is not necessary to
solve the saddle point equations since the normalization
condition for $\rho_k$ can replace this information. 
For the real solution, $s_k$, the requirement  $\prod z_k =1$ forces 
$\lambda^M$ to be real.
We can therefore identify a total of $M$  saddle points
simply related to the real one by overall phases.
These saddle points, arising from the positive real solution to 
(\ref{sequation}) will be termed dominant.
\begin{eqnarray}
z_k^{(p)} &=& e^{2\pi p i/M}z_k^{(0)},\qquad p=0,1\dots M-1\\
{z_k^{(0)}} &=& {{s_k^2\over \rho_k}\over \left(\prod_{k=0}^{M-1} {s_k^2\over \rho_k}\right)^{1/M}}\\
\lambda^{(p)} &=& {\lambda^{(0)}} e^{-2\pi p i/M}\\
{\lambda^{(0)}} &=& M\left(\prod_{k=0}^{M-1} {s_k^2\over \rho_k}\right)^{1/M}
\end{eqnarray}
Note that for the case in which $\rho_k=1$ and the solution
$s_k$ is independent of $k$, these equations simplify dramatically:
${\lambda^{(0)}} = Ms^2$ and the  saddle points are equally spaced
around  the unit circle.

It is tempting to identify each of these $M$ saddle point solutions
with a single dominant saddle point for each of the $M$ integrals $Z^p$.
In that case, the phase factors cancel in the leading contribution
and each $Z^p$ contributes equally
to the total partition function, resulting
in an overall factor of $M$ that 
can be interpreted as arising from the redundancy of cyclic
permutations or equivalently the choice of origin of the
paths. However, the derivation above does not provide an
unambiguous identification of the saddle points with the 
$Z^p$ since the eigenvalues are not globally labeled.
A suitable approach would be to label eigenvalues according to their
order at some reference point, say $z_k=1$ where they are real, 
and then to smoothly deform $z_k$.
This procedure is subject to difficulties since the constraint
$\prod z_k =1$ should be maintained whilst deforming, and
incorrect conclusions arise if one takes
all $z_k = e^{i\theta}$ and varies $\theta=0$ to $\theta=2\pi p/M$.
Some intuition can be gained from explicitly considering 
small $M$ examples and in the appendix the  TSP on two points
is treated. In that case, although the picture is complicated
by the existence of another solution to (\ref{sequation}),
the identification is correct, 
with the dominant contribution (from the positive real solution)
yielding a single saddle point for each of the two eigenvalues.


Henceforth we shall assume that the picture of saddle points given
above is correct. In the next subsection we consider the 
order $1/N$ fluctuations about the saddle point, but first
we deal with the leading contribution from the dominant saddle points. 
The free energy is,
\begin{equation}
-\beta F(\beta,\lbrace \rho_k\rbrace) = \log Z_N =
{N\over M} \sum_{k} \rho_k \log \lambda z_k  + N \log N  
\end{equation}
Where we have used the Stirling approximation for $n_k!$.

The expected path length, $\langle E \rangle$, 
and other quantities such as correlations along the path can be computed
from the partition function by appropriate derivatives.
The expression for the path length can be rewritten 
in terms of $s_k$ itself, but higher moments explicitly involve 
derivatives. 
The final equations determining the leading contribution 
in this coarse regime are:
\begin{eqnarray}
s_k &=& {1\over M}\sum_{k'} e^{-\beta J_{kk'}}{\rho_{k'} \over s_{k'}}\\
\beta F &=&  -\log Z_N = -2N {1\over M} \sum_{k} \rho_k \log s_k - N \log N \\
\langle E \rangle &=& 
{N\over M^2}\sum_{kk'} {\rho_k \rho_{k'} J_{kk'}\over s_k s_{k'}}e^{-\beta J_{kk'}}
\label{eq:hiTdiscrete}
\end{eqnarray}

Notice that the expected path length is always order $N$ in
the limit we have considered. This is indeed expected at finite
temperatures where the number of relevant contributing paths 
grows as a factorial of $N$. To see this we may compute
the entropy as:
\begin{equation}
S(\beta,\lbrace \rho_k\rbrace) 
=
\beta \left(E - F\right)
=
\left(1- \beta {\partial \over \partial \beta}\right)
\left[{2N\over M} \sum_{k} \rho_k \log s_k \right]+ N \log N 
= N S_E +N \log N 
\end{equation}
Where in the final form we have isolated the extensive, $S_E$, from the
non-extensive contributions to the entropy. 
Then by inverting the expression for the path length obtained
from equations (\ref{eq:hiTdiscrete}), the entropy can be 
re-expressed  as a function of path length.
The significance of the entropy in this problem is that the 
number of paths of given length is:
\begin{equation}
{\cal N}(E)
\sim   
e^{S(E)}
= N!\  e^{N S_E}
\label{nofe}
\end{equation}
The factorial growth in the number of paths with the number of cities
characterizes this $N\gg M$ regime and
indicates that  this is not the regime where the optimum path can be found.
To approach that regime, where cities are linked to one of their 
nearest neighbours thus  making the number of paths grow exponentially,
some scaling is necessary. Either the distances $J_{ij}$ can be
scaled so as to preserve constant density of cities as $N\to \infty$,
or equivalently the temperature can be scaled with $N$ \cite{tspps1}.
This kind of scaling would invalidate the simple use of the saddle point
to solve for the partition function as the large parameter 
$N$ would not appear as a simple factor in the exponent.
Therefore the leading order of this coarse model cannot access the optimal paths
of interest in the TSP. However, as we recall below, 
 the approach can be used at finite temperature and
in studying the maximal TSP problem \cite{mtsp}
where no scaling is needed to find the regime of maximal optima.




\subsection{Continuum Geometric Models}

In order to relate the leading order site disordered TSP results above
to the results of our previous functional
approach we take a continuum limit with indices $j$ corresponding to points
${\bf r}_j$ in some $d$ dimensional domain. The distance matrix is defined by
the Euclidean distances between the points, 
$J_{ij} = |{\bf r}_i-{\bf r}_j|$.
Assuming a domain with unit volume,
the appropriate lattice spacing scale is $a= M^{-1/d}$ and
the translations necessary to obtain the continuum limit are
\begin{eqnarray}
\sum_k &\rightarrow& M \int d^dr\\
n_k &\rightarrow& {N\over M} \rho(r_k)
\end{eqnarray}
Where $\rho(r)$ is the continuum density of cities.

In the simplest procedure, the points are laid down on
a regular lattice and the continuum limit is straightforward. 
In this case there is no randomness whatsoever. 
We could also imagine $M$ points laid down at random on the domain,
then some weight factor is needed as we take the continuum version
of the discrete sum. This weight factor appears in exactly the
same places that the density of cities $\rho$ does and 
in the end they are the same. Irrespective of the way the
continuum limit is taken, the formalism only
depends on the density of cities and all information about
the precise location of the sites disappears.

The equations now become,
\begin{eqnarray}
s({\bf r}) &=& \int d^dr e^{-\beta |{\bf r}-{\bf r}'|}
{\rho({\bf r}') \over s({\bf r}')}\\
\beta F &=&  -2N \int d^dr  \rho({\bf r})\log s({\bf r}) - N \log N \\
\langle E \rangle &=& 
N \int d^dr d^dr' {\rho({\bf r})\rho({\bf r}') \over s({\bf r})s({\bf r}')}
|{\bf r}-{\bf r}'|\, e^{-\beta |{\bf r}-{\bf r}'|}
\label{eq:continuum}
\end{eqnarray}
These are the same equations written down in \cite{short} using
a functional formalism in the continuum.
We now briefly summarize some of the results for low dimensional
models based on these equations. The most interesting predictions of
this theory are for the MAX TSP  \cite{mtsp} corresponding to negative $\beta$.


\begin{figure}
\epsfxsize=0.5\hsize \epsfbox{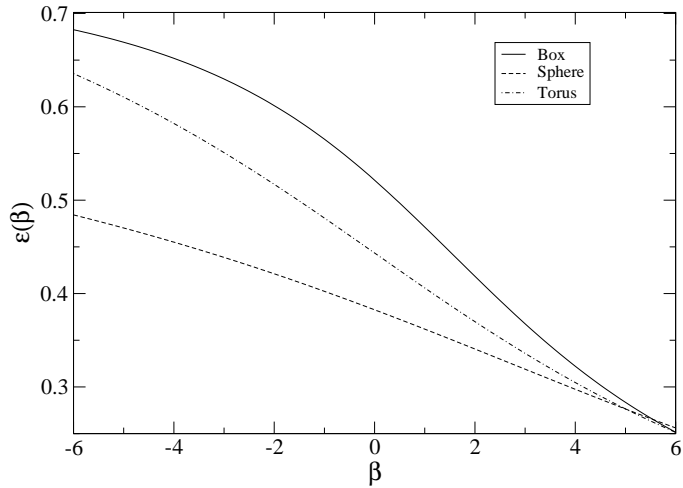}
\caption{Theoretical prediction for the average path length $E$ for 
two dimensional continuum TSP on a box, sphere and torus as a function 
of $\beta$.
Negative $\beta$
corresponds to the MAX TSP.}
\label{fig:2DTSP}
\end{figure}

The formalism presented above yields the correct 
high temperature expansion
as a series in $\beta$, and has also been tested by
comparison with Monte Carlo simulation of the TSP at
finite temperature ~\cite{long}.
Except for the case of closed symmetric domains to be
discussed below, 
we have not found any non-trivial analytic solutions 
of the nonlinear equations
(\ref{eq:continuum}).
For the special closed domains, such as
a disc and torus in two dimensions,
a constant solution exists. 
The significance of this observation is that the annealed
approximation, in which the average over cities is taken
at the level of the partition function Eq. (\ref{eqpart}), is exact for 
these domains. 


For the sphere, the annealed/quenched equations 
have a constant solution with:
\begin{equation}
s^2 = 
{1\over 4\pi} \int_{S_2} d^2r \ e^{-\beta\theta}
={2\pi \over (\beta^2 +4\pi)} (1 + e^{-\sqrt{\pi}\beta/2})
\end{equation}
leading to an expression for the average path length,
\begin{equation}
E/N
= {2\beta \over \beta^2 +4\pi} +  {\sqrt{\pi} \over 2(e^{\sqrt{\pi}\beta/2}+1)}
\label{eqsphere}
\end{equation}
Note that in the limit $\beta \to -\infty$, the path length per link
becomes the  half circumference corresponding, as is the case for all
the closed domains, to the 
maximum distance two points can be apart. 

\medskip
For a torus we find that
the equations yield
\begin{equation}
s^2 = {8 \over \beta^2} \int_0^{\pi/4}
(1 - e^{-\beta/2\cos\theta} - {\beta e^{-\beta/2\cos\theta}\over 2\cos\theta}
)\,d\theta
\label{eqtorus}
\end{equation}
As  $\beta \to -\infty$, the integral can be approximated and
again the path length per link becomes the maximum distance 
two points can be apart, $E/N \to 1/\sqrt{2}$.

\medskip

For most domains, including the traditional square box domain,
the quenched
result is different from the annealed approximation and there
is little hope of a general analytic solution.
In this case, our primary tool is
the iterative numerical solution of Eq. (\ref{eq:continuum})
which is stable and can be solved to any required accuracy.

Figure (\ref{fig:2DTSP}) shows the average path length for 
each of the three two-dimensional domains we have considered.
For the sphere this is given by (\ref{eqsphere}), for the
torus it is based on numerical integration of (\ref{eqtorus})
and for the box we resort to an iterative solution of the
original quenched equations. The accuracy of this iterative technique is
confirmed by reproducing the results for the other domains.
In all these cases we have also performed Monte Carlo simulations
and obtain excellent agreement with the theory.
At large positive $\beta$ the topology starts to become 
unimportant and each domain has average path length $\sim 2N/\beta$.
In dimension $d$ we expect $E \sim d N/\beta$, so this
leading term vanishes at low temperature.

\subsection{Fluctuations}

Despite the inability of the leading term of this approach to 
access minimal length paths, it is interesting to consider the
fluctuations which can contain terms with the correct scaling form.
Gaussian fluctuations about the saddle point yield subleading
corrections in $N$, containing a factor  $det^{-1/2} {\bf H}$, where
${\bf H}$ is the matrix,
\begin{equation}
{\bf H}_{kk'}= -{ \partial^2 \lambda \over \partial\mu_k\partial{\mu_k'}}
=  {\partial v_k^2 \over \partial\mu_{k'}}\vert_{SP}.
\end{equation}
In this coarse regime, the order of the corrections from these fluctuations will
be $1/N$ with respect to the leading $O(N)$  term in the
free energy and will thus correspond to an $N$-independent term.
In one dimension, such an $N$-independent term corresponds
to the scaling of the true TSP optimum.
To proceed explicitly requires additional knowledge of the
distance matrix. We shall consider the special
case of $\rho_k=1$ and a circulant ${\bf B}$ where 
a constant solution $s$ exists and the eigenvectors
are well known to be related to Fourier transforms \cite{circulant}.
This restricted choice of
circulant matrix still covers various interesting 
geometrically based possibilities
and the slightly different case of block circulant matrix
is a simple generalization. Rather than use the general determinant formula
above, we proceed by explicitly expanding about the saddle points.
We assume the picture of $M$ dominant saddles and find that 
besides the phase cancellation at leading order, all
saddle points contribute equally at the level of fluctuations,
and we only need to consider fluctuations about the real
saddle point, which in this case is at $z_k=1$. 
To expand about this saddle point write,
$z_k = e^{i\mu_k}$, where $\mu_k$ is small but may in general be complex.
Then the transfer matrix is given by:
\begin{equation}
{\bf T} = 
\left({\bf I}-{i\over 2} {\bf R}_1 - {1\over 8}{\bf R}^2_1+\dots\right) 
{\bf B}
\left({\bf I}-{i\over 2} {\bf R}_1 - {1\over 8}{\bf R}_1^2+\dots\right)
=  
{\bf T}^{(0)}+{\bf T}^{(1)}+{\bf T}^{(2)}\dots
\end{equation}
Where the matrix ${\bf R}_1$ is diagonal and given by:
\begin{eqnarray}
{\bf R}_1 &=& diag(\dots  \mu_k \dots)
\end{eqnarray}
and
\begin{eqnarray}
{\bf T}^{(0)} &=&   {\bf B}\\
{\bf T}^{(1)} &=& -{i\over 2}
\left( {\bf R}_1{\bf B} + {\bf B}{\bf R}_1\right)\\
{\bf T}^{(2)} &=& -{1\over 8}
\left( {\bf R}_1^2{\bf B} + 2{\bf R}_1{\bf B}{\bf R}_1 
+ {\bf B}{\bf R}_1^2\right)
\end{eqnarray}

The eigenvalue is written in a similar manner as
$\lambda^{(0)} + \lambda^{(1)}
+ \lambda^{(2)} + \dots$
(not to be confused with the labeling of the different eigenvalues
of $\bf T$), and we shall usually drop the superscript 
on the unperturbed $\lambda^{(0)} = \lambda_0 = M s^2$.
The first and second order 
corrections can now be found using the standard techniques of
quantum mechanical perturbation theory (though note that the
matrix ${\bf R}_1$ is not generally Hermitian).
It is convenient to use Dirac bracket notation for the
eigenvectors of the unperturbed ${\bf T}^{(0)}= {\bf B}$.
\begin{equation}
{\bf B}\vert k \rangle = \lambda_k \vert k \rangle
\end{equation}
According to the analysis above, the maximal eigenvector
$\vert 0 \rangle$ has constant components  $v_j = 1/\sqrt{M}$ for
constant density. The choice of a circulant $\bf B$ also
fixes the other eigenvectors, 
$\vert k \rangle$ to have components  $v_j = e^{-2\pi i jk}/\sqrt{M}$.

The first order correction to the eigenvalue vanishes as expected,
\begin{equation}
\lambda^{(1)} = \langle 0\vert {\bf T}^{(1)} \vert 0\rangle
= -{i\lambda_0\over M}\sum_{k=0}^{M-1} {\mu_k} = 0
\end{equation}
At second order, there are two contributions,
\begin{equation}
\lambda^{(2)} = 
 \langle 0\vert {\bf T}^{(2)} \vert 0\rangle
+\sum_{k\ne 0}
{ \langle 0\vert {\bf T}^{(1)} \vert k\rangle
 \langle k\vert {\bf T}^{(1)} \vert 0\rangle
\over \lambda_0 - \lambda_k}
\end{equation}
The expression for the partition function (\ref{coarseZ}) 
involves the following integral over the Gaussian fluctuations:
\begin{equation}
Z_N = M \left[ \left({N\over M}\right)!\right]^M
\lambda_0^N 
\quad
\int  \prod_{k=0}^{M-1}  {d\mu_k\over 2\pi} 
\exp \left({N\lambda^{(2)}\over \lambda_0}\right)
\end{equation}
Where the terms in the exponent arise from expanding the
logarithm and the prefactor $M$ just counts the contribution from
all the equivalent saddle points.
To perform the integrals it is convenient to change
from the $\mu_k$'s to
Fourier transform variables $a_j$ in which the exponent will be diagonal.
\begin{eqnarray}
\mu_k &=& {1\over \sqrt{M}}\sum_{j=0}^{M-1} a_j e^{2\pi i jk/M}\\
a_j &=& {1\over\sqrt{M}}\sum_{k=0}^{M-1} \mu_k e^{-2\pi i jk/M}
\end{eqnarray}
The constant part of $\mu_k$ corresponding to the Fourier mode $a_0$
imposes the constraint which we have implicitly assumed so the 
integration is omitted, and indeed no $a_0$ terms arise in
the integrand.
Using these variables we find,
\begin{eqnarray}
 \langle 0\vert {\bf T}^{(2)} \vert 0\rangle
&=& 
{-1\over 4M}\sum_{j=1}^{M-1} (\lambda_0+\lambda_j)a_ja_{M-j}\\
\sum_{k\ne 0}
{ \langle 0\vert {\bf T}^{(1)} \vert k\rangle
 \langle k\vert {\bf T}^{(1)} \vert 0\rangle
\over \lambda_0 - \lambda_k}
&=&
{-1\over 4M}\sum_{j=1}^{M-1} 
{(\lambda_0+\lambda_j)^2\over(\lambda_0-\lambda_j)}a_ja_{M-j}\\
\lambda^{(2)}
&=& 
{-\lambda_0\over 2M}\sum_{j=1}^{M-1} 
{(\lambda_0+\lambda_j)\over(\lambda_0-\lambda_j)}a_ja_{M-j}
\end{eqnarray}
It is now straightforward to evaluate the integral over the Gaussian
fluctuations since we can take a contour in the direction where
$\mu_k$ is real and therefore $a_j^*=a_{M-j}$.
This corresponds to the imaginary direction in the $z_k$
plane and for other dominant saddles the integration should
be along the tangent to the unit circle.
In this representation the fluctuations are already diagonal
and it is clear that all
the eigenvalues of the fluctuation matrix are
all negative.
\begin{equation}
\beta F =
-2N\log s + {1\over 2} \sum_{j\ne0} 
\log\left({\lambda_0+\lambda_j\over\lambda_0-\lambda_j}\right)
- N \log N
\label{flucteqn}
\end{equation}
The final fluctuation part is a simple expression in terms of the
eigenvalues of the coupling matrix $\bf B$.
Below we consider some particular examples with circulant $\bf B$ in this
$N\gg M$ regime, the fully connected 
and geometrically motivated  problems where expressions for the
eigenvalues are available and the computation of the fluctuations
can be taken further.


\subsection{Fully Connected Graph}

Consider the regular fully connected graph 
$J_{ij} = J_N(1-\delta_{ij})+J_D\delta_{ij}$
where all sites are connected by equal length links and 
we have allowed a diagonal term.
We write $e^{-\beta J_D} = d$ and $e^{-\beta J_N} = n$ to obtain the
circulant form:
\begin{equation}
{\bf B} = \left(\begin{array}{ccccc}
    d&n&n&\cdots&n\\
    n&d&n&\cdots&n\\
    n&n&d&\cdots&n\\
    \vdots&\vdots&\vdots& & \vdots\\
    n&n&n&\cdots&d
\end{array}\right)
\end{equation}

For a uniform distribution of cities $\rho_k = 1$, the equations
(\ref{sequation})
have a uniform solution $s_k = s$ where,
\begin{equation}
s^2 = {1\over M} \left(d + (M-1)n \right).
\end{equation}
Saddle points are equally spaced about the unit circle
and the eigenvalue $\lambda^{(0)} = Ms^2$.

Based on the dominant saddle points associated with this solution,
the leading contribution to the
expected path length is,
\begin{equation}
{\langle E \rangle \over N} = 
{d J_D  + (M-1)n J_N \over
d + (M-1)n}
\end{equation}
At low temperature, depending on the relative size of the 
diagonal and neighbour distances, we obtain 
$\langle E \rangle = N J_D$ or $\langle E \rangle = N J_N$.
These results correspond to two regimes in which paths
either prefer to revisit the same site or move to a
neighbouring site. 

As explained at the beginning of this section,
we hope that the fluctuation term may yield interesting information
in the limit $J_N > J_D$ when $\beta \to \infty$, {\it ie} $n/d \to 0$.
Fluctuations around the dominant saddle points can be 
evaluated as described above since this problem has a circulant
cost matrix.
The eigenvalues are very
degenerate and besides the maximal eigenvalue given above
all others take the value $d-n$. 
Using the expression (\ref{flucteqn}) we find:
\begin{equation}
\beta F = -N \log\left(d + (M-1) n\right) 
+{M-1\over 2}\log\left(2d/n + (M-2)\right)
+(N-{M-1\over 2})\log M -N \log N 
\end{equation}
So in the limit  $n/d \to 0$ the path length is:
\begin{equation}
{\langle E \rangle } = 
\left[N - {M-1\over 2}\right] J_D + {M-1\over 2} J_N
\end{equation}
In this limit the fluctuations about the saddle point fail to give the expected 
optimum: $(N-M) J_D + M J_N$.
The reasons for the failure concern branch points approaching the
saddle points and possibly other (non-dominant) solutions 
becoming important. Indeed other solutions for $s_k$
besides the symmetric solution exist.
The details of the failure are most clearly exposed in the
simplest $M=2$ example given in the appendix.

\subsection{1-D lattice}

Several models corresponding to a one dimensional lattice with nearest 
neighbour couplings can be imagined depending on the boundary
conditions and symmetry of the couplings. On a ring with uniform distribution 
of cities $\rho_k = 1$, we may simply allow nearest neighbour connections,
$J_{ij} = J_N\delta_{i,j+1}+J_N\delta_{i,j-1}+J_D\delta_{ij}$, or 
consider the possibility of longer links,
$J_{ij} = J_N|i-j|+J_D\delta_{ij}$ or
$J_{ij}=J_N(N-|i-j|)+J_D\delta_{ij}$,
according to which is the closest
way around the ring. In each case we have allowed
a diagonal term, $J_{ii}=J_D$.
The generalizations to problems on a line with fixed boundaries
do not have a circulant form, but the difference is only a
surface effect which corresponds to ignoring terms
of order $e^{-M\beta/2}$.

In the nearest neighbour case we have the circulant form:
\begin{equation}
{\bf B} = \left(\begin{array}{ccccc}
    d&n&0&\cdots&n\\
    n&d&n&\cdots&0\\
    0&n&d&\cdots&0\\
    \vdots&\vdots&\vdots& & \vdots\\
    n&0&0&\cdots&d
\end{array}\right)
\end{equation}
The equations (\ref{sequation})
then have a uniform solution $s_k = s$ where,
\begin{equation}
s^2 = {1\over M} \left(d + 2n \right)
\end{equation}
The leading term in the path length is then:
\begin{equation}
{\langle E \rangle \over N}= 
{d J_D + 2 n J_N\over
d + 2 n}
\end{equation}
At low temperature, the picture is similar to the fully
connected case. Depending on the relative size of the 
diagonal and neighbour distances, we obtain 
$\langle E \rangle = N J_D$ or $\langle E \rangle = N J_N$,
corresponding to regimes in which paths
either prefer to revisit the same site or move to a
neighbouring site.

The eigenvalues at the typical saddle point $z_k = 1$ are:
\begin{equation}
\lambda_k = d + 2n\cos(2\pi k/M)
\label{oneDeigenvalues}
\end{equation}
So using (\ref{flucteqn}),
the free energy may be written:
\begin{equation}
\beta F =
-N \log (d + 2 n) +
{1\over 2} \sum_{k=1}^{M-1} 
\log\left({d+2 n \cos^2\pi k/M\over n\sin^2\pi k/M}\right)
+(N-1/2)\log M - \log N!
\end{equation}
For  a large number of sites, $M$, 
the sum can be approximated as:
\begin{eqnarray}
\sum_{k=1}^{M-1} 
\log\left({d+2 n \cos^2\pi k/M\over n\sin^2\pi k/M}\right)
&\to&
(M-1)\log 2+
{M\over 2\pi}\int_0^{2\pi}d\omega\,
\log\left( {d/n+1+\cos\omega\over1-\cos\omega}\right)\\
&=&M\log 2 
+M\log\left( {d\over n}+ 1+\sqrt{\left({d\over n}\right)^2 + 2{d\over n}}\right)
\end{eqnarray}

The corrections to the path length can then be read off, and in
the limit $n/d \to 0$ we have:
\begin{equation}
E =
\left[ N-{M\over 2} \right]J_D+{M\over 2} J_N
\label{ringlowT}
\end{equation}
The finite correction, were it
not for the factor of 2, would be exactly what we expect
for the ordinary TSP. The approach fails again for technical
problems with the saddle point approach in this limit as
is most clearly exposed in the appendix. An additional
non-dominant solution for $s_k$ is found in this case also.

These results can be generalized to 
one dimensional models with long range interactions and
regular periodic lattices in higher dimensions with block circulant
$\bf B$. However, in view of the difficulties at the simplest
level, we do not pursue this route.


\section{Fully Occupied $N=M$ Regime}

With the occupation numbers set to $n_k=1$ and $N=M$
we recover the usual formulation of the TSP with a
given cost matrix $J_{ij}$ that was discussed in
section II.
\begin{equation}
Z_N(\beta)   =
\oint  \prod_{k=0}^{N-1}  {dz_k\over 2\pi i  } 
\exp \left(
N \log \lambda\right)
=
\int_{-\pi}^{\pi}  \prod_{k=0}^{N-1}  {d\mu_k\over 2\pi } 
\exp \left(\sum_{k'=1}^N i\mu_{k'}
+N \log \lambda_{max}\right).
\end{equation}
For a regular lattice there is no randomness in this regime
and the TSP optimum is a trivial problem.
However, it is difficult to recover the trivial result
from the approach based on the expression above.
For example, the 1 dimensional lattice can only be solved 
explicitly for small $N$.
We investigate this case in some detail to illustrate
that the saddle point approach in principle yields the
scaling that would allow a low temperature limit to
access minimal path lengths, but that other reasons
invalidate the approach.
Moreover, we  hope that results on regular
lattices might form the basis of an expansion towards
more dilute lattices.



The saddle point method cannot be applied 
directly since the factor $N$ does not multiply all terms in the 
exponent, but consider the following (unitary) change of variables
\begin{eqnarray}
\mu_k &=& {1\over \sqrt{N}}\sum_{j=0}^{N-1} a_j e^{2\pi i jk/N}\\
a_j &=& {1\over\sqrt{N}}\sum_{k=0}^{N-1} \mu_k e^{-2\pi i jk/N}
\end{eqnarray}
This is a 1-dimensional discrete Fourier transform, but
note that it is independent of any possible
dimension  underlying the $J_{ij}$'s and holds for arbitrary
cost matrices.
Unitary normalization has been chosen so 
the Jacobean is one and we obtain,
\begin{equation}
Z_N(\beta)   =
\int  \prod_{j=0}^{N-1}  {da_j\over 2\pi } 
\exp \left(i\sqrt{N}a_0
+N\log \lambda\right).
\end{equation}
Now, by considering the structure of the general transfer 
matrix, we find that the  $a_0$ dependence factors out 
so the eigenvalues are of the form
$\lambda(a_0,a_1\dots,a_{N-1})= e^{-a_0/\sqrt{N}}
\lambda'(a_1\dots,a_{N-1})$. 
Therefore all  $a_0$ dependence cancels in the action
as we expect since this mode simply imposes the implicit
constraint  $\sum n_k = N$ that we have satisfied by setting $n_k=1$.
Integration over this mode should be omitted to obtain:
\begin{equation}
Z_N(\beta)  =
\int  \prod_{j=1}^{N-1}  {da_j\over 2\pi } 
\exp \left(N
\log \lambda'(a_1,a_2\dots a_{N-1})\right).
\end{equation}
A saddle point technique can again be attempted, but we must take care
that the eigenvalue itself does not have any residual $N$ dependence.
In the case of the $N\gg M$ regime, this was straightforward as the
matrix was $M\times M$, here it is $N\times N$, but 
for local $J_{ij}$ where the number of nearest neighbours
does not grow with $N$, the eigenvalue can be $N$ independent.

The stationary conditions from varying $a_k$, $k=1,2\dots N-1$ are,
\begin{equation}
\sum_{j=0}^{N-1} e^{2\pi i jk/N}v_j^2 = 0
\end{equation}
So, when complemented with the normalization condition
on $v_k$ we have,
\begin{equation}
v_k^2 = 1/N
\label{spcondition}
\end{equation}
resembling the saddle point condition obtained in the $N\gg M$ regime.
As in the $N\gg M$ regime there are spurious signed solutions of the
saddle point equations that do not give a distinct saddle point
from the solution with  $v_k = +1/\sqrt{N}$.
Proceeding through similar steps to that regime and
writing 
$s_i = \sqrt{\lambda'} \exp(\mu_i/2 - \sqrt{N} a_0/2)$,
we obtain the final equations,
\begin{eqnarray}
s_k &=& \sum_{k'=0}^{N-1} {e^{-\beta J_{kk'}} \over s_{k'}}
\label{seqn}\\
\beta F &=&  -\log Z_N = -2\sum_{k}  \log s_k \\
\langle E \rangle &=& 
\sum_{jj'} { J_{jj'}e^{-\beta J_{jj'}}\over s_j s_{j'}}
\label{eq:NeqMdiscrete}
\end{eqnarray}
Note that  no $1/N$ factors appear in front of the sums, 
however, for local cost matrices at sufficiently low temperature
only neighbours will contribute to
the sums, and 
we find an $N$ independent expression for $s_i$. 
Such a solution suggests that the saddle point is valid
as it has no untoward $N$ dependence.
Moreover, this constitutes a correct low temperature scaling solution 
since the nearest neighbour distance is fixed as $N$ increases and the
size of the domain increases keeping the density of
cities constant. We therefore expect that this approach might
access sufficiently low temperatures to provide information 
about the optima.

To investigate further we take into account the
fluctuations about the saddle point, and
following the discussion for the coarse regime 
we consider $\bf B$ to be a symmetric circulant matrix.
For these special distance matrices
the saddle point equations have a constant solution $z_k=z$,
the transfer matrix becomes ${\bf T}= z^{-1} {\bf B}$ and
all eigenvalues are directly related to those of $\bf B$.
Using the scaling symmetry of the $a_0$ mode we can choose
$z=1$. In fact there are a set of $N$ saddles with $z$ any $N$'th
root of unity, but they all contribute equally to
the integral and we only need study the one at $z=1$.
Determining the Hessian
matrix becomes an exercise in second order perturbation theory
that when aided by the structure of the matrix yields the following
quadratic correction  to the exponent:
\begin{equation}
{1\over 2}\sum_{k=1}^{N-1} {\lambda_0+\lambda_k\over\lambda_0-\lambda_k}
|a_k|^2
\end{equation}
Where the eigenvalues $\lambda_k$ are of the unperturbed transfer matrix, 
which at this constant saddle is simply $\bf B$ ($\lambda_0$ is the maximal
eigenvalue corresponding to the eigenvector with all 1's).
Note that the zero mode has dropped out as expected, and that in the
$a_k$ variables the Hessian is already diagonal.
With some care about the direction of integration over the saddle,
the quadratic correction to the free energy is:
\begin{equation}
\beta F =  -\log Z_N = -N \log \lambda_0 +
{1\over 2}\sum_{k=1}^{N-1} 
\log\left({\lambda_0+\lambda_k\over\lambda_0-\lambda_k}\right)
\end{equation}
Note that there are $N$ terms in the eigenvalue sum and typically
each term in the sum contributes.
For example, for the nearest
neighbour one dimensional ring
the eigenvalues are given by (\ref{oneDeigenvalues})
and the sum is order $N$.
The fluctuation correction is of the
same order as the leading term indicating a failure of the
saddle point approach since we can anticipate similar magnitude  
corrections at cubic, quartic etc levels.

\section{Dilute Lattice Regime: $N < M$}

The dilute regime is the most interesting from the site disorder 
perspective. We consider a regular lattice with sites either
occupied or not, $n_k = 0,1$.  As the system becomes very
dilute, the discrete nature of the underlying lattice can be ignored
and we recover the traditional TSP. Indeed, numerical TSP
optimization codes use integer arithmetic following this principle.
Here we would be content to investigate the leading correction
to the path length for small dilution as 
the numerical simulations~\cite{chakrabati}
seem to suggest a fairly linear
dependence of energy against probability of site occupation.

For the first time in this work, it is necessary to
explicitly average over the disorder, in this case the
occupation numbers. This must be a quenched average and 
for this kind of site-disordered problem 
it is natural to use a formalism originally due to
Morita~\cite{morita} and recently championed by Kuhn~\cite{kuhn}.
Essentially the method consists of taking an annealed
average, but inserting constraints to fix the distribution of
the occupation numbers at its quenched value.
In practice this is achieved through an approximation
in which only moments of the distribution are fixed.
As a first step we would only consider the first two moments from the hierarchy
corresponding to the constraints:
\begin{eqnarray}
\sum_{k=0}^{M-1} n_k &=& N\\
\sum_{<kk'>} n_kn_{k'} &=& d{N^2\over M}
\end{eqnarray}
The first constraint is straightforward and is incorporated simply by 
including the integration over the $a_0$ mode that was dropped
in earlier sections since the set of $n_k$'s chosen already
obeyed the constraint. The second constraint involves
a sum over nearest neighbours and
prevents the occupied sites from clustering together thus
minimizing path lengths.
\begin{equation}
Z_N(\beta) \approx \sum_{n_k = 0,1}
\delta\left(\sum_{<kk'>} n_kn_{k'} - d{N^2\over M}\right)
\int \prod_{k=0}^{M-1} d\mu_k e^{i\sum n_k \mu_k}
\sum_p \lambda_p^N
\end{equation}
This approach relies on a better understanding of the 
unperturbed, $N=M$, lattice and takes
us away from the aim of this paper, so we defer 
further study for the future.

\section{Conclusions}

We have introduced  a well defined discrete problem:
the site disordered TSP with $N$ cities on $M$ sites, and have studied
it using a saddle point approach. 

\begin{itemize}
\item In the coarse, $N\gg M$ limit, we obtain a discrete version
of the continuum theory we developed previously using a functional
approach. The discrete approach shows that $M$
saddle points contribute to the result rather than the single
one identified in the functional work, but that 
(at least for constant density of cities) they are
all equivalent. The effect of fluctuations can
also be computed using the discrete approach. However,
in the limit where there is a diagonal distance much 
shorter than neighbour distances 
and this correction might give information about the minimal
path length in one dimensional problems, 
the saddle point approach suffers technical
difficulties and gives incorrect results. 
The nature of the difficulties were exposed in the $M=2$ example
as being associated with additional saddle points
becoming relevant and saddle points approaching branch
cuts. 

\item The saddle point approach to the  ordinary $N=M$ TSP
has the correct scaling
to be able to access the interesting low temperature regime
even at the leading order.
However, this fails to give correct results 
since the saddle is shallow and $O(N)$ modes all contribute 
to the fluctuations about the saddle generating a correction that
is of the same order as the leading term thus invalidating the approach.

\item We have discussed the dilute model and speculated on 
an approach based on an approximation originally due to Morita.

\item We expect that it would be possible to use the saddle point method
to analyze models that
are intermediate between the $N\gg M$ regime and the $N=M$ case
with city densities that vary as $N^\alpha$ for some power $0<\alpha\le 1$. 
The  quadratic fluctuations in these models will be of order
$N^{1-\alpha}$ and generally allow an expansion in $N^{-\alpha}$,
but we anticipate the same problems with the low temperature
limit.

\end{itemize}

Of course a saddle point approach is not the only way to proceed
and one interesting aspect of the formalism we have presented is that 
it describes the path weights in terms of particular terms in
a generating function thus resembling approaches to other
combinatorial optimization problems. This approach
can certainly be used to enumerate for small $N$.

\vskip 0,5 truecm

\noindent{\bf Acknowledgment:} D.L. would like to
acknowledge a Discipline Hopping Award from the EPSRC
during which much of this work was done.

\appendix

\section{Explicit  Calculation for $M=2$}

The general derivation of the coarse regime given in the text
did not label the eigenvalues globally and concentrated on
the real positive solution of the saddle point equation.
Explicit computations for particular
small values of $M$ illuminate the role of other eigenvalues and saddle
points.

\begin{figure}
\epsfxsize=0.5\hsize \epsfbox{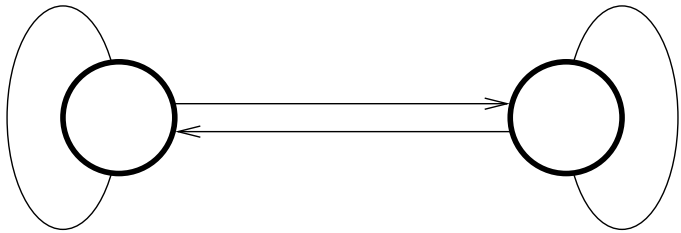}
\caption{TSP on two sites.}
\label{Mtwopic}
\end{figure} 

\begin{figure}
\epsfxsize=0.4\hsize \epsfbox{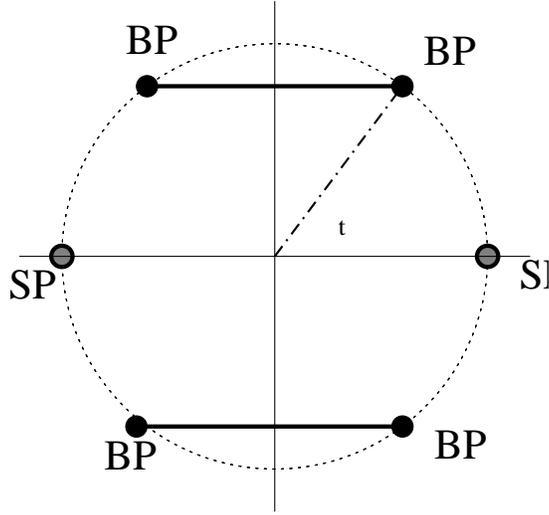}
\caption{Complex plane for TSP on two sites. For $\rho_-=0$ and 
in the region where
$n<d$, where it is cheaper to revisit a site than to move to the other
site. The branch points (BP), cuts and two saddle points (SP) are shown.
The angle $t$ is given by $\sin t = n/d$.}
\label{M2zplane}
\end{figure}

Consider the case $M=2$ (figure \ref{Mtwopic}) where 
a diagonal entry is allowed in the
distance matrix as otherwise the path is completely determined.
This model can be reinterpreted as a 1-dimensional Ising model
on a ring with $N$ sites subject to the constraint that states must
have zero overall magnetization. The analysis of the Ising 
model provides additional intuition, but a constrained transfer
matrix approach repeats essentially the same technique as in the text.
Writing $d=e^{-\beta J_D}$ for the diagonal term and 
$n=e^{-\beta J_N}$ for the neighbour term, the parameters of
the system are then $d/n$ and $\rho_{-} = (\rho_0-\rho_1)/2$
using the terminology of section \ref{coarse}.
The partition function
is the sum of the contributions from each of the two eigenvalues
$Z= n_0!n_1!(Z_++Z_-)$ where,
\begin{equation}
Z_\pm   =
\oint  {dz_0\over 2\pi i z_0  } {dz_1\over 2\pi i z_1 } \
z_0^{n_0}z_1^{n_1}\
\lambda_\pm^N
\end{equation}
and $\lambda_\pm$ are the two eigenvalues of the transfer matrix,
\begin{equation}
{\bf T}= {1\over z_0z_1}
\left(\begin{array}{cc}
    d z_1& n\sqrt{z_0z_1}\\
    n\sqrt{z_0z_1}&dz_0
\end{array}\right)
\label{M2Tmata}
\end{equation}
In this form, one of the integration variables simply imposes
the constraint $n_0+n_1 =N$ that we will generally make implicit.
The constraint is therefore redundant and if we proceeded in this
way the two saddle point equations would be linearly dependent
and only set $z_0$ and $z_1$ proportional to each other. 
While this approach merely leads to a volume factor, it is clearer 
to remove the constraint from the start. We therefore insert
the factor $2\pi i\delta(z_0z_1 -1)$ 
(the $2\pi i$ is the Jacobean factor) into the partition function
to obtain a one dimensional contour integral:
\begin{equation}
Z_\pm   = 
\oint  {dz\over 2\pi i z } 
z^{N\rho_-/2}
\lambda_\pm^N
\end{equation}
Where $\lambda_\pm$ are now the two eigenvalues of the transfer matrix,
\begin{equation}
{\bf T}= 
\left(\begin{array}{cc}
    d/z& n\\
    n&dz
\end{array}\right)
\label{M2Tmatb}
\end{equation}
That is:
\begin{equation}
\lambda_\pm = {1\over 2}\left( {d}(z+{1\over z}) 
\pm \sqrt{d^2 (z-{1\over z})^2 +4n^2}\right)
\end{equation}
In this form it is possible to keep track of the same eigenvalues,
labeled by $\pm$ or their relative magnitude at $z=1$, 
and not swap between them as $z$ varies around the integration contour.
Note that this has involved a choice of sign in moving between
(\ref{M2Tmata}) and (\ref{M2Tmatb}), for example,
consider $z=z_0=z_1=-1$ (obeying the constraint) 
in each version of the transfer matrix.
The branch points and saddle points in the complex plane are shown
in figure \ref{M2zplane}. For $n<d$ it is possible to choose a contour
on the unit circle, but for $n>d$, the contour must be deformed 
away from the branch cuts.

For certain values of the parameters the integrals can be computed
directly. For $d=0$ or $n=0$, a contribution is only obtained
for certain choices of $n_0$, $n_1$. We find $d=0$ is only 
possible if $n_0=n_1=N/2$ in which case $Z=2 (N/2)!^2 n^N$.
For $n=0$ then one of $n_0$ or $n_1$ must vanish. These restrictions
follow from the integrals but match the expectations from the 
original geometric problem since the choice of connections must 
make it possible for paths 
to visit nodes the required number of times.
For $n=d$, one eigenvalue vanishes and we obtain 
$Z= N! d^N$ irrespective of $n_0$, $n_1$ since there are
a total of $N!$ paths each of equal weight.
The limit of $n\ll d$ is interesting  as the corrections to
the leading term provide information about the real (not coarse)
TSP. In this limit, the diagonal path is short and is always 
preferred over the neighbour path.
By expanding we obtain $Z= n_0!n_1! N n^2 d^{N-2}$ indicating 
that there are $N$ paths that only flip once between nodes
and this matches the combinatorial approach.

Now we take it that neither $d$ nor $n$ vanishes and use the 
saddle point approach. The stationarity equation is:
\begin{equation}
{\rho_-\over  z} + {1\over \lambda_\pm} 
{d\lambda_\pm\over d z} = 0 
\end{equation}
which can be solved to find two saddle points for each eigenvalue.
These saddle points depend on the sign of $\rho_-$, but without loss
of generality we can always take the 
sign to be positive. 
\begin{eqnarray}
z^2_{sp} = 1+2g^2 \mp \sqrt{g^2 + g^4}\\
g^2 = {n^2 \rho_-^2\over d^2 (1-\rho_-^2)}
\end{eqnarray}
The upper and lower signs correspond to the two different eigenvalues,
while the two separate saddle points for each eigenvalue arise from the
sign of $z_{sp}$.

In general the saddle points are located at
different points depending on the eigenvalue in question.
Now consider the symmetric case of $\rho_-=0$ where $g^2=0$,
where the location of the saddle points at $z_{sp} = \pm 1$,
is the same for each eigenvalue. For $Z_+$ involving the
eigenvalue $\lambda_+$,
the value of the eigenvalue at the saddle points 
$z_{sp} = \pm 1$ is $(\pm d+n)$ with eigenvector $(1,1)$
at each saddle. For the other integral, $Z_-$, involving 
eigenvalue $\lambda_-$,
the value of the eigenvalue at the saddle points 
$z_{sp} = \pm 1$ is $(\pm d-n)$ with eigenvector $(1,-1)$
at each saddle point. 
It is already apparent that for each integral, one
of the saddle points dominates the other (since we 
assumed $d \ne 0$), for $Z_+$ the dominant saddle is at
$z_{sp}=+1$, whereas for $Z_-$ it is $z_{sp}=-1$.

These explicit computations match the expectations from the
general development in the text. The dominant saddle points
come from the real positive solution of equation (\ref{sequation}),
and the  subdominant saddle points
arise from a distinct solution of the equation (\ref{sequation}).
The eigenvector associated with the dominant saddle point
depends on which integral, $Z_\pm$, is considered.

The asymptotic value of integrals for $Z_\pm$ can be obtained
from the dominant  saddle point in each case. By expanding about 
the dominant saddle we find that the contour should pass over it
in the imaginary direction, and the contribution from fluctuations
can be evaluated. Each of $Z_\pm$ yields the same contribution since
$N$ must be even for the symmetric case under consideration.
Overall we obtain:
\begin{equation}
Z   = (N/2)!^2 \sqrt{2 n\over\pi N d} (d+n)^N 
\approx N! \sqrt{n\over d} \left({d+n\over 2}\right)^N 
\end{equation}
This expression reproduces the exact result at $n=d$, but
the limits $d\to 0$ or $n\to 0$ do not reproduce the 
known results.
In such limits the simple saddle point calculation fails
for technical reasons.
The subdominant saddles would be expected to contribute 
(in which case the contour
over them should be in the real direction) and moreover the
branch points can approach the saddle points  or the cuts can
force the contour to be deformed.
A more careful approach via a uniform asymptotic expansion
\cite{saddle} may give better results.

In summary, the coarse TSP on two points allows us to carefully
track each eigenvalue and find that
multiple saddles can occur for each eigenvalue. 
One saddle point dominates each integral associated with a certain eigenvalue. 
In the full partition function each of these eigenvalues contribute equally. 
The eigenvector corresponding to the dominant saddle depends on
the eigenvalue being tracked.
Finally, the limit of removing elements of the distance matrix
by making them large can be delicate in the saddle point approach.
In particular we are unable to recover the true TSP as a 
subleading contribution to the $n\ll d$ limit from the saddle point.

Already for the $M=3$ generalization, the coarse TSP on 3 sites,
the algebra to allow an eigenvalue to be tracked 
becomes excessive even for $d=0$ and
with the aid of a computer.  Additional isolated solutions 
of the saddle point equations 
in the text for $s_i$ can be found provided $d\ne0$.


\pagestyle{plain}


\baselineskip =18pt
\begin{thebibliography}{0}
\bibitem{tspps2}{M. M\'ezard and G. Parisi, Europhys. Lett. {\bf 2}, 913 
(1986).}
\bibitem{tspcav}{W. Krauth and M. M\'ezard,  Europhys. Lett. {\bf 8}, 213 
(1989).}
\bibitem{short}{D.S.~Dean, D.~Lancaster and S.N.~Majumdar, J. Stat. Mech.
L01001 (2005).}    
\bibitem{long}{D.S.~Dean, D.~Lancaster and S.N.~Majumdar, Phys. Rev. E {\bf 72}
026125 (2005).}
\bibitem{dsjohnson}{D.S.~Johnson and L.A.~McGeoch, 
in G. Gutin and, A. P. Punnen (Eds),
{\it The Traveling Salesman Problem and its Variations} 
(Combinatorial Optimization Series),  Kluwer, Boston (2002).}
\bibitem{chakrabati}{M.~Ghosh, S.S.~Manna and B.K.~Chakrabarti,
J.Phys {\bf A} 21, 1483 (1988);
A.~Chakraborti, and  B.K.~Chakrabarti, Eur. Phys.J. {\bf B} 16, 677 (2000).}
\bibitem{circulant}{Philip J. Davis. {\it Circulant Matrices}. 
Chelsea Pub Co; 2nd edition (1994).}
\bibitem{wallpaper}{R.S. Garfinkel, Oper. Res. {\bf 25},
 741 (1977).}
\bibitem{BrendanMcKay}{
B. D. McKay and N. C. Wormald, European J. Combin., {\bf 11} (1990) 565-580;
B. D. McKay, Combinatorica, {\bf 10} (1990) 367-377.}
\bibitem{Zee}{M. M\'ezard, G. Parisi, and A. Zee, Nucl. Phys. {\bf B559}, 
689 (1999);
A. Zee and I. Affleck, J. Phys: Condens. Matter {\bf 12}, 8863 (2000).}
\bibitem{HeldKorp}{R.M.~Karp and J.M.~Steele,
in E.L.~Lawler, J.K.~Lenstra, A.H.G.~Rinnooy Kan and D.B.~Shmoys (Eds),
{\it The Traveling Salesman Problem}, Wiley, (1985).}
\bibitem{saddle}
{N.G. De Bruijn, {\it Asymptotic Methods in Analysis}
North-Holland, Amsterdam (1958);
F.W.J. Olver, {\it Asymptotics and Special Functions}
Academic Press, New York (1974);
R. Wong, {\it Asymptotic Approximations of Integrals}
Academic Press, San Diego (1989).}
\bibitem{tspps1}{J. Vannimenus and M. M\'ezard, J. Physique Lett.
{\bf 45} L1145 (1984).}
\bibitem{mtsp}{A. Barvinok, E. Kh. Gimadi and A.I. Serdyukov, 
in G Gutin and, A. P. Punnen (Eds),
{\it The Traveling Salesman Problem and its Variations} 
(Combinatorial Optimization Series),  Kluwer, Boston (2002).}
\bibitem{morita}{T.~Morita J. Math. Phys. {\bf 5}, 1401, (1964).}
\bibitem{kuhn}{R.~K\"{u}hn, Z. Phys. {\bf 100}, 231 (1996).}
\end{thebibliography}
\end{document}